\documentclass[preprint,showpacs,amsmath]{revtex4}
\usepackage{amssymb}
\usepackage{graphicx,epsfig,subfigure,dsfont,amssymb,amsthm,amsfonts,amsbsy,mathrsfs,amscd}
\def\qed{\leavevmode\unskip\penalty9999 \hbox{}\nobreak\hfill
     \quad\hbox{\leavevmode  \hbox to.77778em{%
               \hfil\vrule   \vbox to.675em%
               {\hrule width.6em\vfil\hrule}\vrule\hfil}}
     \par\vskip3pt}

\input amssym.def
\begin{document}

\begin{center}
\textbf{Coherence evolution in two-qubit system going through amplitude damping channel}
\end{center}

\begin{center}
{Ming-Jing Zhao$^{1}$, Teng Ma$^{2}$, Yuquan Ma$^{1}$}

\begin{minipage}{5in}
\small $~^{1}$ {School of Science, Beijing
Information Science and Technology University, 100192, Beijing,
China}

\small $~^{2}$ {State Key Laboratory of Low-Dimensional Quantum Physics and Department of Physics, Tsinghua University, Beijing 100084, China}

\end{minipage}
\end{center}

\textbf{Abstract}\ \
In this paper, we analyze the evolution of quantum coherence in a two-qubit system going through the amplitude damping channel. After they had gone through this channel many times, we analyze the systems with respect to the coherence of their output states. When only one subsystem goes through the channel, frozen coherence occurs if and only if this subsystem is incoherent and an auxiliary condition is satisfied for the other subsystem. When two subsystems go through this quantum channel, quantum coherence can be frozen if and only if the two subsystems are both incoherent. We also investigate the evolution of coherence for  maximally incoherent-coherent states and derive an equation for the output states after one or two subsystems have gone through the amplitude damping channel.

\textbf{Keywords}\ \ Quantum coherence, Amplitude damping channel, Frozen coherence

\textbf{PACS}\ \ 03.65.Ud, 03.67.Mn

\section{Introduction}
Coherence, the superposition of state in quantum theory \cite{J. Aberg,T. Theurer}, constitutes a powerful resource in quantum metrology \cite{V. Giovannetti}, entanglement creation \cite{J. K. Asboth,M. Gao}, and biological systems \cite{E. Collini,N. Lambert,J. Cai,J. M. Cai,E. J.OReilly}.
Since the inception of quantum mechanics, many approaches have been proposed for incorporating this important
feature \cite{A. Streltsov-rev}.
In Ref. \cite{T. Baumgratz}, the authors introduced a rigorous framework for the quantification of coherence and identified some intuitive and
computable measures of coherence including, for example relative entropy coherence and $l_1$ norm coherence. A state's  relative entropy coherence is defined as the difference of  von Neumann entropy between a density matrix and a diagonal matrix formed by its diagonal elements. The $l_1$ norm coherence depends on the absolute value
of the off-diagonal  elements of the density matrix.
The authors in Ref. \cite{A. Winter} introduced coherence distillation and coherence cost in operational way by focusing
on the optimal rate of performance of certain tasks, which revealed that the distillable coherence equals to the relative entropy coherence.
Coherence can also be converted to entanglement via incoherent operations and the authors in Ref. \cite{A. Streltsov} introduced geometric coherence on this basis.
In Ref. \cite{L. H. Shao}, the authors showed  that coherence fidelity in general does not  satisfy the monotonicity requirement of a coherence measure, and they reported that trace norm coherence can act as a measure of coherence for qubits.
Moreover, the authors in Ref. \cite{S. Rana} reported that trace norm coherence is also a strong monotone for all qubit states and X states. In addition,  coherence can also be quantified via convex roof construction \cite{X. Yuan}.

However, quantum coherence is typically recognized as a fragile feature, and the disappearance of coherence in open quantum systems exposed to environmental noise is commonly referred to as decoherence \cite{A. Streltsov-rev}.
As quantum resources, it is important to identify the conditions under which quantum coherence does not deteriorate during open evolution. As such, the concept of frozen coherence was proposed and the authors in Ref. \cite{T. R. Bromley} investigated the dynamical conditions under which coherence is totally
unaffected by quantum noise. With just one qubit system, no nontrival condition exists such that $l_1$ norm coherence and relative entropy coherence are simultaneously frozen under any quantum channel.
For a high dimensional quantum system, all measures of coherence are frozen in an initial state in a strictly incoherent channel
if and only if the relative entropy coherence is frozen \cite{X. D. Yu}. In Ref. \cite{M. L. Hu}, the authors proved a factorization relation for $l_1$ norm coherence
under some special conditions and obtained a condition for frozen coherence.

In this paper we investigate the evolution of quantum coherence in the two-qubit system. Here we focus on $l_1$ norm coherence and analyze the dynamics of quantum coherence in the amplitude damping channel. We find that if one subsystem goes through the amplitude damping channel many times,  frozen coherence can appear if and only if this subsystem is incoherent and the other subsystem fulfills an additional requirement. If two subsystems go through this quantum channel, then frozen coherence can appear if and only if the two subsystems are both incoherent. As an example, we analyze the evolution of coherence of maximally incoherent-coherent states in the amplitude damping channel. We also derive
an equation for the output states occurring after one or two subsystems go through the amplitude damping channel.

\section{Coherence evolution under amplitude damping channel in two-qubit system}

\subsection{Theory preliminary}

\emph{Quantum states}~~
First, we fix computational basis $\{|i\rangle\}$ as the reference basis in each local subsystem, which we then use throughout this paper. Using this local reference basis, quantum states can be classified with respect to the type of quantum coherence. In the bipartite system, a quantum state is called incoherent-coherent if it can be written as $\rho=\sum_{i} p_i |i\rangle\langle i|\otimes \rho_i$. Similarly, it is called coherent-incoherent if it can be written as $\rho=\sum_{i} p_i \rho_i \otimes|i\rangle\langle i| $ \cite{E. Chitambar}. Incoherent-coherent and coherent-incoherent states are incoherent in one subsystem. Their coherence are the average of the coherent parts, $C(\rho)=\sum_i p_i C(\rho_i)$.
In fact, for the incoherent-coherent state, one postulate for a measure of coherence \cite{A. Streltsov-rev} requires that $C(\rho)$ be nonincreasing on average under selective incoherent operations, so
$C(\rho)\geq\sum_i p_i C(|i\rangle\langle i|\otimes \rho_i)=\sum_i p_i C(\rho_i)$, for which we choose the incoherent operation for the local projective measurements $\{|i\rangle\langle i|\}$ on the first subsystem. Another postulate for a measure of coherence \cite{A. Streltsov-rev} requires that $C(\rho)$ be a convex function of density matrices, which implies $C(\rho)\leq \sum_i p_i C(|i\rangle\langle i|\otimes \rho_i) =\sum_i p_i C(\rho_i)$. Therefore, the coherence of incoherent-coherent and coherent-incoherent states is the average of the coherent parts.
They are maximally coherent if and only if all components $\rho_i$ in the coherent part are maximally coherent.
If a quantum state is incoherent in two subsystems,  it is called incoherent and is written as $\rho=\sum_{ij} p_{ij} |i\rangle\langle i|\otimes |j\rangle\langle j|$.
The classification of quantum states with respect to quantum coherence parallels the classification of quantum correlation \cite{S. Luo2008}, the former  relevant to the reference basis and the latter independent of reference basis.

\emph{Quantum coherence}~~
A very intuitive quantification of coherence
would relate to the off-diagonal elements of
the quantum state being considered. Therefore, it is desirable to quantify the
coherence by a function that depends on the off-diagonal
elements \cite{T. Baumgratz}. For the quantum state $\rho=\sum_{ij} \rho_{ij}|i\rangle\langle j|$, the $l_1$ norm coherence
under the given reference basis $\{|i\rangle\langle i|\}$ is given by
\begin{eqnarray}
C(\rho)=\sum_{i\neq j} |\rho_{ij}|.
\end{eqnarray}
For $d$-dimensional quantum state $\rho$, its coherence is bounded by
\begin{equation}
0\leq C(\rho)\leq d-1,
\end{equation}
since $C(\rho)=2 \sum_{i< j} |\rho_{ij}|
= 2 \sum_{i< j} \sqrt{|\rho_{ij}|^2}
\leq 2 \sum_{i< j} \sqrt{\rho_{ii}\rho_{jj}}
\leq  \sum_{i< j} (\rho_{ii}+\rho_{jj})
=d-1$.
It is maximally coherent if and only if
\begin{equation}
\rho_{ii}=|\rho_{ij}|=\frac{1}{d}
\end{equation}
for $i\neq j$, $i,j=0,\cdots, d-1$.

\emph{Amplitude damping channel}~~ The amplitude damping channel is a quantum operation that describes the energy dissipation-effects due to the loss of energy in a quantum system. Suppose we have a single optical mode $a_0|0\rangle+a_1|1\rangle$ with $|a_0|^2+|a_1|^2=1$, i.e., a superposition of zero or one photons. The scattering of photon from this mode can be modeled by inserting  a beam splitter on the path of the photon. This beam splitter allows the photon to couple with another single optical mode, $|0\rangle$, according to the unitary transformation $B=\exp[\theta(a^\dagger b- ab^\dagger)]$, where $a, a^\dagger$ and $b, b^\dagger$ are annihilation and creation operators for photons in the two modes respectively. Assuming that the environment starts out with no photons, the output state after beamsplitter is simply $B|0\rangle(a_0|0\rangle+a_1|1\rangle)=a_0|00\rangle+a_1(\cos\theta|01\rangle+\sin\theta|10\rangle)$. By tracing over the environment, we obtain the following amplitude damping operation \cite{M. A. Nielsen}
\begin{eqnarray}
{\cal{E}}(\rho)=E_0\rho E_0^\dagger + E_1\rho E_1^\dagger,
\end{eqnarray}
where
\begin{eqnarray*}
E_0=\left(
\begin{array}{cc}
1& 0\\
0 & \sqrt{1-\gamma}\\
\end{array}
\right),
E_1=\left(
\begin{array}{cc}
0& \sqrt{\gamma}\\
0 & 0\\
\end{array}
\right),
\end{eqnarray*}
$0 \leq\gamma\leq 1$.

\subsection{Coherence evolution under amplitude damping channel in two-qubit system}

Next, we consider a quantum state in the two-qubit system
\begin{eqnarray}\label{two qubit state}
\rho=\left(
\begin{array}{cccc}
a_{11}& a_{12} & a_{13} & a_{14}\\
a_{21}& a_{22} & a_{23} & a_{24}\\
a_{31}& a_{32} & a_{33} & a_{34}\\
a_{41}& a_{42} & a_{43} & a_{44}
\end{array}
\right),
\end{eqnarray}
with $\rho\geq 0$, $a_{ji}=a_{ij}^*$ for $i\neq j$, $\sum_{i=1}^4 a_{ii}=1$.
The coherence is
\begin{eqnarray}
C(\rho)=2\sum_{i< j} |a_{ij}|.
\end{eqnarray}

In the first case, we let the first subsystem goe through the amplitude damping channel, and the output state $\rho_{L}^{(1)}$ for $\rho$ is as follows:
\begin{eqnarray}
\rho_{L}^{(1)}=E_0\otimes I \rho E_0^\dagger\otimes I + E_1\otimes I \rho E_1^\dagger\otimes I,
\end{eqnarray}
where the subscript $L$ indicates that the first subsystem goes through the amplitude damping channel.
If the first subsystem goes through this channel twice, the output state $\rho_{L}^{(2)}$ is as follows:
\begin{eqnarray}
\rho_{L}^{(2)}=E_0\otimes I \rho_{L}^{(1)} E_0^\dagger\otimes I + E_1\otimes I \rho_{L}^{(1)} E_1^\dagger\otimes I.
\end{eqnarray}
If the first subsystem goes through the amplitude damping channel $n$ times, then the output state $\rho_{L}^{(n)}$ is as follows:
\begin{eqnarray}
\rho_{L}^{(n)}=E_0\otimes I \rho_{L}^{(n-1)} E_0^\dagger\otimes I + E_1\otimes I \rho_{L}^{(n-1)} E_1^\dagger\otimes I,
\end{eqnarray}
which can be rewritten as
\begin{eqnarray}
\rho_{L}^{(n)}=\sum_{i_1,i_2,\cdots, i_n=0,1}  E_{i_1i_2\cdots i_n} \otimes I \rho E_{i_1i_2\cdots i_n}^\dagger\otimes I
\end{eqnarray}
with $E_{i_1i_2\cdots i_n}=E_{i_1}E_{i_2}\cdots E_{i_n}$. Due to the properties of operators $E_0$ and $E_1$ in the amplitude damping channel,
\begin{eqnarray}
E_1^2=0, \ \ E_0E_1=E_1,\ \ E_1E_0=\sqrt{1-\gamma}E_1,
\end{eqnarray}
$\rho_{L}^{(n)}$ is reduced to the sum of $n+1$ terms as follows:
\begin{eqnarray}
\rho_{L}^{(n)}=E_0^n \otimes I \rho (E_0^n)^\dagger\otimes I+\sum_{i=0}^{n-1} E_1 E_0^{n-i-1} \otimes I \rho (E_1 E_0^{n-i-1})^\dagger \otimes I.
\end{eqnarray}

By straightforward calculation, we obtain the following:
\begin{eqnarray}\label{left rho}
\rho_{L}^{(n)}=\left(
\begin{array}{cccc}
a_{11}+a_{33}[1-(1-\gamma)^n]& a_{12}+a_{34}[1-(1-\gamma)^n] & a_{13}(1-\gamma)^{\frac{n}{2}} & a_{14}(1-\gamma)^{\frac{n}{2}}\\
a_{21}+a_{43}[1-(1-\gamma)^n]& a_{22}+a_{44}[1-(1-\gamma)^n] & a_{23}(1-\gamma)^{\frac{n}{2}} & a_{24}(1-\gamma)^{\frac{n}{2}}\\
a_{31}(1-\gamma)^{\frac{n}{2}}& a_{32}(1-\gamma)^{\frac{n}{2}} & a_{33}(1-\gamma)^{n} & a_{34}(1-\gamma)^{n}\\
a_{41}(1-\gamma)^{\frac{n}{2}}& a_{42}(1-\gamma)^{\frac{n}{2}} & a_{43}(1-\gamma)^{n} & a_{44}(1-\gamma)^{n}
\end{array}
\right).
\end{eqnarray}
The coherence is as follows:
\begin{eqnarray}\begin{aligned}
C(\rho_{L}^{(n)})&=2\{|a_{12}+a_{34}[1-(1-\gamma)^n]|\\
&+(1-\gamma)^{\frac{n}{2}}(|a_{13}|+|a_{14}|+|a_{23}|+|a_{24}|+|a_{34}|(1-\gamma)^{\frac{n}{2}})\},
\end{aligned}
\end{eqnarray}
which is smaller than or equal to $C(\rho)$, $C(\rho_{L}^{(n)})\leq C(\rho)$. They coincide if and only if $a_{13}=a_{14}=a_{23}=a_{24}=0$ and $|a_{12}+a_{34}[1-(1-\gamma)^n]|=|a_{12}|+|a_{34}|[1-(1-\gamma)^n]$. This means that the input state is as follows:
\begin{eqnarray}\label{special inco-co}
\rho=|0\rangle\langle0|\otimes \left(
\begin{array}{cc}
a_{11}& a_{12}\\
a_{21}& a_{22}
\end{array}
\right)+
|1\rangle\langle1|\otimes \left(
\begin{array}{cc}
a_{33}& a_{34}\\
a_{43}& a_{44}
\end{array}
\right),
\end{eqnarray}
with $a_{12}$ and $a_{34}$ having the same argument, $a_{12}=a_{21}^*$, $a_{34}=a_{43}^*$, and $\sum_{i=1}^4 a_{ii}=1$. So if the first subsystem goes through the amplitude damping channel, frozen coherence occurs only for incoherent-coherent state in the form of Eq. (\ref{special inco-co}) with $a_{12}$ and $a_{34}$ having the same argument for the second subsystem.

If we consider the case in which $n$ tends to infinite,  $C(\rho_{L}^{(n)})$ tends to $2|a_{12}+a_{34}|$. So if the first subsystem goes through the amplitude damping channel many times, the coherence can not disappear for quantum states if $a_{12}+a_{34}\neq 0$. This implies that the coherence is robust under the influence of this channel.

Similarly, when the second subsystem goes through the amplitude damping channel many times, the output state is as follows:
\begin{eqnarray}
\rho_{R}^{(n)}=\sum_{i_1,i_2,\cdots, i_n=0,1} I\otimes E_{i_1i_2\cdots i_n}  \rho I\otimes E_{i_1i_2\cdots i_n}^\dagger,
\end{eqnarray}
where the subscript $R$ indicates that the second subsystem goes through this channel.
For the quantum state $\rho$ in Eq. (\ref{two qubit state}), the output state $\rho_{R}^{(n)}$ is as follows:
\begin{eqnarray}\label{right rho}
\rho_{R}^{(n)}=\left(
\begin{array}{cccc}
a_{11}+a_{22}[1-(1-\gamma)^n] & a_{12}(1-\gamma)^{\frac{n}{2}} & a_{13}+a_{24}[1-(1-\gamma)^n] & a_{14}(1-\gamma)^{\frac{n}{2}} \\
a_{21}(1-\gamma)^{\frac{n}{2}} & a_{22}(1-\gamma)^{n} & a_{23}(1-\gamma)^{\frac{n}{2}} & a_{24}(1-\gamma)^{n}\\
a_{31}+a_{42}[1-(1-\gamma)^n] & a_{32}(1-\gamma)^{\frac{n}{2}} & a_{33}+a_{44}[1-(1-\gamma)^n] & a_{34}(1-\gamma)^{\frac{n}{2}}\\
a_{41}(1-\gamma)^{\frac{n}{2}} & a_{42}(1-\gamma)^{n} & a_{43}(1-\gamma)^{\frac{n}{2}} & a_{44}(1-\gamma)^{n}
\end{array}
\right).
\end{eqnarray}
Its coherence is as follows:
\begin{eqnarray}\begin{aligned}
C(\rho_{R}^{(n)})&=2\{|a_{13}+a_{24}[1-(1-\gamma)^n]|\\
&+(1-\gamma)^{\frac{n}{2}}(|a_{12}|+|a_{14}|+|a_{23}|+|a_{34}|+|a_{24}|(1-\gamma)^{\frac{n}{2}})\},
\end{aligned}\end{eqnarray}
which is smaller than or equal to the coherence of the input state, $C(\rho_{R}^{(n)})\leq C(\rho)$. These coincide if and only if $a_{12}=a_{14}=a_{23}=a_{34}=0$ and $|a_{13}+a_{24}[1-(1-\gamma)^n]|=|a_{13}|+|a_{24}|[1-(1-\gamma)^n]$, which means that the input state is as follows:
\begin{eqnarray}\label{special co-inco}
\rho=\left(
\begin{array}{cc}
a_{11}& a_{13}\\
a_{31}& a_{33}
\end{array}
\right)\otimes |0\rangle\langle0|+
\left(
\begin{array}{cc}
a_{22}& a_{24}\\
a_{42}& a_{44}
\end{array}
\right)\otimes |1\rangle\langle1|,
\end{eqnarray}
with $a_{13}$ and $a_{24}$ having the same argument, $a_{13}=a_{31}^*$, $a_{24}=a_{42}^*$, and  $\sum_{i=1}^4 a_{ii}=1$.
So if the second subsystem goes through the amplitude damping channel, frozen coherence occurs only for coherent-incoherent state in the form of Eq. (\ref{special co-inco}) with $a_{13}$ and $a_{24}$ having the same argument for the first subsystem.

Furthermore, if two subsystems both go through the amplitude damping channel, the output state is as follows:
\begin{eqnarray}
\rho^{(n)}=\sum_{i_1,i_2,\cdots, i_n,j_1,j_2,\cdots, j_n=0,1} E_{i_1i_2\cdots i_n}\otimes E_{j_1j_2\cdots j_n}  \rho E_{i_1i_2\cdots i_n}^\dagger\otimes E_{j_1j_2\cdots j_n}^\dagger.
\end{eqnarray}
By straightforward calculation, we have the following:
{\small
\begin{multline}
\rho^{(n)}=\left(
\begin{array}{cccc}
a_{11}+a_{33}[1-(1-\gamma)^n] & \{a_{12}+a_{34}[1-(1-\gamma)^n]\}(1-\gamma)^{\frac{n}{2}}& a_{13}(1-\gamma)^{\frac{n}{2}} & a_{14}(1-\gamma)^{n} \\
\{a_{21}+a_{43}[1-(1-\gamma)^n]\}(1-\gamma)^{\frac{n}{2}} &  \{a_{22}+a_{44}[1-(1-\gamma)^n]\}(1-\gamma)^{n} &a_{23}(1-\gamma)^{n} & a_{24}(1-\gamma)^{\frac{3n}{2}}\\
a_{31}(1-\gamma)^{\frac{n}{2}} & a_{32}(1-\gamma)^{n} &a_{33}(1-\gamma)^{n} & a_{34}(1-\gamma)^{\frac{3n}{2}}\\
a_{41}(1-\gamma)^{n} & a_{42}(1-\gamma)^{\frac{3n}{2}}&a_{43}(1-\gamma)^{\frac{3n}{2}} & a_{44}(1-\gamma)^{2n}
\end{array}
\right)\\
+\left(
\begin{array}{cccc}
\{a_{22}+a_{44}[1-(1-\gamma)^n]\}[1-(1-\gamma)^n] & 0 & a_{24}(1-\gamma)^{\frac{n}{2}}[1-(1-\gamma)^n] & 0 \\
0 &  0 & 0 & 0\\
a_{42}(1-\gamma)^{\frac{n}{2}}[1-(1-\gamma)^n] & 0 & a_{44}(1-\gamma)^{{n}}[1-(1-\gamma)^n] & 0\\
0 & 0 & 0 & 0
\end{array}
\right).
\end{multline}
}

The coherence of $\rho^{(n)}$ is $C(\rho^{(n)})=2\{|a_{12}+a_{34}[1-(1-\gamma)^n]|+ |a_{13}+a_{24}[1-(1-\gamma)^n]|+(|a_{14}|+|a_{23}|)(1-\gamma)^{\frac{n}{2}}+
(|a_{24}|+|a_{34}|)(1-\gamma)^{n}\}(1-\gamma)^{\frac{n}{2}}$.
The coherence is frozen after two subsystems go through the amplitude damping channel if and only if $\rho$ is incoherent, which means that the coherence in the coherent state can not be frozen under this assumption.

Since, in this scenario, only the coherence in coherent-incoherent and incoherent-coherent states can be frozen forever, we now examine them more closely. For example,
for the following two-qubit incoherent-coherent state
\begin{eqnarray}\label{incoherent quantum state}
\rho=p_0 |0\rangle\langle0|\otimes \rho_0 + p_1|1\rangle\langle1|\otimes \rho_1 ,
\end{eqnarray}
where $\rho_0=\left(
\begin{array}{cc}
a_{11}& a_{12}\\
a_{21} & a_{22}
\end{array}
\right)$, $\rho_1=\left(
\begin{array}{cc}
a_{33}& a_{34}\\
a_{43} & a_{44}
\end{array}
\right)$,
$a_{12}=a_{21}^*$, $a_{34}=a_{43}^*$, and $a_{11}+a_{22}=a_{33}+a_{44}=1$, its coherence is $C(\rho)=2(p_0|a_{12}|+p_1|a_{34}|)$.
By the positivity of $\rho_0$ and $\rho_1$, we know $|a_{12}|,|a_{34}|\leq \frac{1}{2}$. Therefore, $0\leq C(\rho)\leq 1$. Obviously, $C(\rho)$ reaches its minimum of zero, if $\rho$ is incoherent. $C(\rho)$ reaches its maximum if and only if both $\rho_0$ and $\rho_1$ are maximally coherent states, which are in the form of $\frac{1}{2}\left(
\begin{array}{cc}
1& e^{i\theta}\\
e^{-i\theta} & 1
\end{array}
\right)$ and pure states.

Now we consider the maximally coherent states in Eq. (\ref{incoherent quantum state}), which we refer to as a maximally incoherent-coherent state and
study the evolution of coherence through the amplitude damping channel.
First, it is easy to determine that $C(\rho_{R}^{(n)})=(1-\gamma)^{\frac{n}{2}}$ which is independent of the quantum states themselves. Coherence $C(\rho_{R}^{(n)})$ decreases strictly as parameter $\gamma$ increases. After further calculation we obtain the following equation
\begin{equation}\label{equation}
C(\rho^{(n)})=C(\rho_{L}^{(n)})C(\rho_{R}^{(n)}),
\end{equation}
which reveals that $C(\rho^{(n)})$ is proportional to $C(\rho_{L}^{(n)})$ since $C(\rho_{R}^{(n)})$ is uniquely determined by the quantum channel. This demonstrates that the evolution of coherence after two subsystems go through the amplitude damping channel is determined by the evolution of coherence when the first subsystem goes through this channel.

Next we analyze the coherence of the maximally incoherent-coherent states when the first subsystem goes through the amplitude damping channel.
If $\rho_0$ and $\rho_1$ are the same maximally coherent state $|+\rangle=\frac{1}{\sqrt{2}}(|0\rangle+|1\rangle)$, then
\begin{eqnarray}
\rho_{m_1}= (p_0|0\rangle\langle0|+p_1|1\rangle\langle1|) \otimes |+\rangle\langle+|.
\end{eqnarray}
Note that the reference basis is computational basis, from which is easy to obtain $C(\rho_{m_1,L}^{(n)})=1$. So if the first subsystem goes through the amplitude damping channel, the coherence can be frozen forever. If $\rho_0$ and $\rho_1$ are the orthogonal maximally coherent pure states $|+\rangle=\frac{1}{\sqrt{2}}(|0\rangle+|1\rangle)$ and $|-\rangle=\frac{1}{\sqrt{2}}(|0\rangle-|1\rangle)$, then
\begin{eqnarray}
\rho_{m_2}=p_0 |0\rangle\langle0| \otimes |+\rangle\langle+|+p_1 |1\rangle\langle1| \otimes |-\rangle\langle-|
\end{eqnarray}
and $C(\rho_{m_2,L}^{(n)})=|p_0-p_1[1-(1-\gamma)^n]|+p_1({1-\gamma})^n$. The coherence can not be frozen for the nontrival case $\gamma\neq 0,1$. When $n$ tends to infinity, the coherence reaches $|p_0-p_1|$. For $p_0\neq p_1$, the coherence can not disappear forever and is robust under the influence of amplitude damping channel. If $\rho_0$ and $\rho_1$ are non-orthogonal maximally coherent pure states, for example, $|+\rangle=\frac{1}{\sqrt{2}}(|0\rangle+|1\rangle)$ and $|r\rangle=\frac{1}{\sqrt{2}}(|0\rangle+i|1\rangle)$, then
\begin{eqnarray}
\rho_{m_3}=p_0|0\rangle\langle0|\otimes |+\rangle\langle+| + p_1 |1\rangle\langle1| \otimes |r\rangle\langle r|.
\end{eqnarray}
$C(\rho_{m_3,L}^{(n)})=\sqrt{p_0^2+p_1^2[1-(1-\gamma)^n]^2}+p_1({1-\gamma})^n$. Neither can the coherence be frozen for the nontrival case $\gamma\neq 0,1$. When $n$ approaches infinity, the coherence reaches $\sqrt{p_0^2+p_1^2}$, which is larger than $|p_0-p_1|$. Hence the coherence in $\rho_{m_3}$ is more robust than the coherence in  $\rho_{m_2}$ under the influence of the amplitude damping channel. In FIG. 1, we show plots of the coherence evolution of $C(\rho_{m_2,L}^{(2)})$ and $C(\rho_{m_3,L}^{(2)})$, in which we can see the rate of change of coherence $C(\rho_{m_2,L}^{(2)})$ and $C(\rho_{m_3,L}^{(2)})$ becomes larger as parameter $\gamma$ becomes larger.

\begin{center}
\begin{figure}[!h]\label{fig}
\resizebox{8cm}{!}{\includegraphics{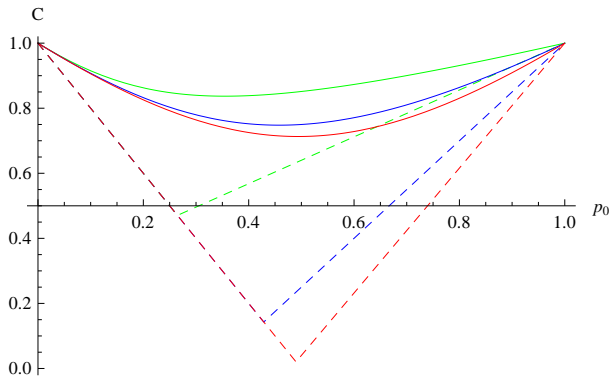}}\caption{(Color online) Plot the coherence evolution for $\rho_{m_2}$ and $\rho_{m_3}$ when the first subsystem goes through the amplitude damping channel twice. The green, blue, and red lines represent the coherence evolution through the amplitude damping channel for $\gamma=0.2, 0.5$, and 0.8 respectively. The solid line represents the coherence of $\rho_{m_3,L}^{(2)}$ and the dashed line the coherence for $\rho_{m_2,L}^{(2)}$.}
\end{figure}
\end{center}

\subsection{Discussions}

The amplitude damping channel is a quantum operation, and, more specially, an incoherent quantum operation. Its operators $K_0$ and $K_1$ map incoherent states into incoherent states. It appears that amplitude damping channel cannot create coherence, and if a quantum state goes through this quantum channel, then its coherence will decrease. It is plausible to assert that in the two-qubit system, if the first subsystem goes through this channel, then the quantum coherence can be frozen for all incoherent-coherent states, since we do nothing to the second subsystem. But our above analysis shows this to be false. Not all coherence in the incoherent-coherent state can be frozen when the first subsystem goes through this channel, i.e., part of this coherence does not deteriorate.

Another quantum operation is the phase damping channel, which describes the loss of quantum information without loss of energy \cite{M. A. Nielsen}. The energy eigenstates of a quantum system do not change as a function of time, but rather accumulate a phase that is proportional to the eigenvalue. Its operators can be represented as $\tilde{K}_0=\left(
\begin{array}{cc}
1& 0\\
0 & \sqrt{1-\lambda}
\end{array}
\right)$ and $\tilde{K_1}=\left(\begin{array}{cc}
0& 0\\
0 & \sqrt{\lambda}
\end{array}
\right)$.
The phase damping channel is also an incoherent quantum operation. Therefore, it cannot increase coherence when a quantum state goes through it. Moreover, because of the commutativity of operators $\tilde{K}_0$ and $\tilde{K}_1$, we derive that for the two-qubit system, when one subsystem goes through this channel, frozen coherence appears if and only if this subsystem is incoherent. Similarly if two subsystems go through this channel, frozen coherence happens if and only if both subsystems are incoherent. By contrast, the amplitude damping channel requires more condition for frozen coherence. Since all ccoherence in incoherent-coherent states can be frozen when the first subsystem goes through the phase damping channel, and only some of coherence can be frozen when going through the amplitude damping channel, more conditions are required for the second subsystem.

Physically, the dynamics in which an atom emits a photon spontaneously, a spin system at high temperature approaches equilibrium with its environment, and the state of a photon in an interferometer or cavity when it is subject to scattering and attenuation are all characterized as amplitude damping channel \cite{M. A. Nielsen}. For example, in the process in which a two level atom coupled with a vacuum undergoes spontaneous emission, the parameter $\gamma$ in the amplitude damping channel is expressed as $1-\exp(-2\Gamma t)$, where $\Gamma$ is a constant and $t$ is time. In this model, repeatedly sending the state into the amplitude damping channel is equivalent to letting the state remain a longer period of time. However, for the process in which a harmonic oscillator interacts with an environment through the Hamiltonian $H=\chi (a^{\dagger} b + b^{\dagger} a)$, the parameter $\gamma$ in the amplitude damping channel is expressed as $\gamma=1-\cos^2(\chi t)$, which denotes the probability of loosing a single quantum of energy. For this process, repeatedly sending quantum state into the amplitude damping channel is now not equivalent to letting it remain for a longer period of time. In short, it is certain that parameter $\gamma$ generally depends on time, but going through quantum channel many times is still a discrete time evolution.

Another matter we want to clarify is that frozen coherence differs from the concept of a ``decoherence free subspace" \cite{D. A. Lidar}. Frozen coherence is the retaining of coherence subjected to an external environment, whereas a decoherence free subspace is a subspace that is invariant under non unitary dynamics. It is the carrier of quantum information and is completely safe from the influence of an environment \cite{M. Demianowicz}.

\section{Conclusions}

In this paper we have analyzed the evolution of quantum coherence in the two-qubit system through the amplitude damping channel. Using calculations, we have analyzed the coherence of the output states of one or two subsystems going through this channel many times. We have found that if one subsystem goes through this quantum channel,  frozen coherence occurs if and only if this subsystem is incoherent and an auxiliary requirement is satisfied for the other subsystem. If two subsystems undergo this quantum channel, frozen coherence occurs if and only if the two subsystems are both incoherent.  We have also analyzed the evolution of coherence for maximally incoherent-coherent states and derived an equation for the output states after one or two subsystems have gone through the amplitude damping channel.

\bigskip
Acknowledgments:  This work is supported by the NSF of China under
Grant Nos. 11401032 and 11404023.

\end{document}